\documentclass[aps,prd,twocolumn,groupedaddress,superscriptaddress,floatfix,showpacs,showkeys]{revtex4}

\usepackage[utf8]{inputenc}
\usepackage[T1]{fontenc}
\usepackage{slashed}
\usepackage{times}

\usepackage{amssymb,amsfonts,amsmath,amsthm}
\usepackage{dsfont,bbm}
\usepackage{dcolumn}

\usepackage{graphicx}
\usepackage[caption=false]{subfig}

\begin{document}

\title{Color entanglement for $\gamma$-jet production in polarized pA collisions}

\author{Andreas~Sch\"afer}
\affiliation{\normalsize\it Institut f\"ur Theoretische
Physik,Universit\"at Regensburg, Regensburg, Germany}

\author{Jian~Zhou}
\affiliation{\normalsize\it Institut f\"ur Theoretische
Physik,Universit\"at Regensburg, Regensburg, Germany}

\begin{abstract}
A more reliable treatment of transverse momentum dependent physics
and in particular transverse single spin asymmetries is urgently
required, e.g. for polarized pA physics including novel effects
like color entanglement. We argue that the measurement of
azimuthal angular correlations of photons and jets produced in pA collisions
provides a direct access to the novel gluon distribution $G_4$ that
enters into many such processes.
\end{abstract}

\pacs      {}

\maketitle

\section{Introduction}

While collinear QCD is by now well understood, the new frontier is
a rigorous treatment of transverse momentum dependent (TMD)
physics. If the detected transverse momenta are not very large
(e.g. several tens of  GeV) many novel, highly non-trivial effects
are relevant which could, e.g., influence the interpretation of heavy ion data
(the $A$ dependence of these effects is basically unknown, making the comparison
of $p+p$, $p+A$ and $A+A$ data for $k_{\perp}$ dependent quantities
highly problematic). One of the tasks of the proposed EIC accelerator
\cite{Boer:2011fh,Accardi:2012qut} is the investigation of this physics
and of the associated differences between $e+p$ and $e+A$ collisions.

There exist many open questions, starting from such fundamental
issues as TMD factorization, see, e.g.
~\cite{Collins:1981uk,Ji:2004wu,GarciaEchevarria:2011rb}. TMD parton
distributions are expressed as matrix elements of nonlocal
combinations of quark or gluon fields which are connected by gauge
links. To take their effects into account a generalized version of
TMD factorization (GTMD)~\cite{Bomhof:2004aw} was proposed
containing process dependent gauge links and thus a modified concept
of universality. However, it was found that in hadron-hadron
reactions, color entanglement leads to additional contributions
which cannot be factorized even within GTMD
factorization~\cite{Rogers:2010dm}. If these effects are large and
cannot be reliably and quantitatively described by theory the
corresponding experimental $k_{\perp}$-dependent data cannot be
interpreted within QCD, severely limiting their value. On the other
hand, the presence of such effects can be seen not as nuisance, but
as chance to explore the nontrivial interplay of color flow in local non-Abelian gauge
theories. Therefore, the discovery of color entanglement effects has
stimulated a lot of theoretical interest, see, e.g.
~\cite{Buffing:2011mj,Buffing:2012sz,Buffing:2013dxa,Schafer:2014zea}.
In this contribution we argue that transverse single spin
asymmetries (SSA) in  $\gamma$-jet production in polarized pA
collisions is a promising channel to do so.

The Sivers effect giving rise to the SSA in this process is particularly sensitive to color entanglement
as its existence relies on the gauge link. We take into account one extra gluon exchange on the
proton side, while the gluon re-scattering is resummed to all orders on the nucleus side
by means of Wilson lines. This approximation is justified by the fact that the gluon number
density in a nucleus is much larger than in a proton. We will
further argue below that higher order gluon exchange
with the remnant part of the proton is suppressed by a factor
$\Lambda_{QCD}/Q_s$ in the semi-hard region, where $Q_s$ is the saturation momentum.

The spin dependent observables in  pA collisions
(e.g. the Sivers asymmetry) are generally affected by saturation.
Therefore, measuring these observables might be a promising approach to
firmly experimentally establish saturation.
 Recent developments in this direction include
 the calculation of quark/gluon Boer-Mulders functions~\cite{Metz:2011wb}
 and the quark Sivers function~\cite{Kovchegov:2013cva} in the quasi-classical
McLerran Venugopalan (MV) model~\cite{McLerran:1993ni},
 the derivation of small x evolution equation for the gluon Boer-Mulders function~\cite{Dominguez:2011br},
 the determination of the asymptotic behavior of SSAs
at small x~\cite{Schafer:2013opa,Zhou:2013gsa},
and the investigation of spin asymmetries in pA collisions
beyond the Eikonal approximation~\cite{Altinoluk:2014oxa}.
SSAs in various processes in polarized pA collisions also have been studied within  different
frameworks~\cite{Boer:2006rj,Kang:2011ni,Kovchegov:2012ga,Kang:2012vm}.
More recently, a hybrid approach was formulated to compute the SSA for inclusive direct photon production in
polarized pA collisions~\cite{Schafer:2014zea}.

In this letter, we study the SSA for photon and jet production in polarized pA collisions
by closely following the method
presented in Ref.~\cite{Blaizot:2004wv,Schafer:2014zea}.
The result can be decomposed into two parts, one of which is
related to the normal dipole type TMD gluon distribution and an additional one
related to the new gluon distribution function $G_4(x,k_{\perp})$ generated by color
entanglement. We will show that the transverse momentum dependence of the
differential cross section in the semi-hard region reads,
\begin{eqnarray}
\frac{d \sigma^{p^{\uparrow} A \to \gamma q + X}}{d P.S.}
 \approx \sum_q H_{Born}
 \left \{
x f_q(x)  \, x_g' G_{DP}(x_g',q_{\perp})\right .
\nonumber \\
- \varepsilon_\perp^{ij}S_\perp^i q_{\perp}^j x T_{F,q}(x,x)
\frac{N_c^2+1}{N_c^2-1} \frac{\partial}{\partial q_\perp^2}x_g' G_{DP}(x_g',q_{\perp})
\nonumber \\
+\left. \varepsilon_\perp^{ij}S_\perp^i q_{\perp}^j x T_{F,q}(x,x)
\frac{2N_c^2}{N_c^2-1} \frac{\partial}{\partial q_\perp^2}x_g' G_{4}(x_g',q_{\perp})
\right \}
\label{eq_1}
\end{eqnarray}
where $f_q(x)$ and $G_{DP}(x_g, q_\perp)$ are the integrated quark PDF in the proton and
the dipole type gluon TMD distribution in the nucleus.
$\varepsilon_\perp^{ij}$ is a short-hand notation for the transverse epsilon
tensor $\varepsilon_\perp^{-+ij}$ defined with the convention $\varepsilon_\perp^{-+12}=1$.
$T_{F,q}(x,x)$ is the well known ETQS function~\cite{Efremov:1981sh,Qiu:1991pp}.
$d P.S.=dy_1 dy_2 d^2l_{\gamma \perp} d^2 l_{q\perp}$, where $y_1$,
 $y_2$ are the rapidities of the two outgoing particles, and $l_{\gamma \perp}$, $l_{q\perp}$
 are  the transverse momenta
of the
produced photon and quark, respectively.
$q_\perp$ is their sum  $q_\perp=l_{\gamma\perp}+l_{q\perp}$.
The hard coefficient is given by,
\begin{eqnarray}
H_{Born} &=& \frac{\alpha_s \alpha_{em} e_q^2}{N_c \hat s^2}
\bigg( -\frac{\hat{s}}{\hat{u}} - \frac{\hat{u}}{\hat{s}}
 \bigg)
\end{eqnarray}
The first two terms in Eq.(\ref{eq_1}) are already obtained in GTMD factorization
while the last one describes the novel color entanglement effect.
 Experimentally, nothing is known about the $q_{\perp}^2$ slope of $G_4$ and it is
unclear if this contribution to the SSA is sizable. Thus, it is very important that this SSA is measured at RHIC
 to find out whether $G_4$ is relevant or not for this observable. If it is, color entanglement
effects could very well be important in general for $q_\perp$ dependent hadronic reactions,
which would greatly complicate the interpretation of existing and future experimental data.

In the following we will present a few details of the calculation.

\section{$\gamma$-jet  SSA in GTMD factorization}
We  consider the kinematic region where the transverse momenta of the produced photon and jet
are much larger than their transverse momentum imbalance  which is often
referred to as the correlation limit.
The dominant partonic channel contributing in the forward region of the proton is:
\begin{equation}
q_p(xP+p_\perp)+ g_A(x_g' \bar P+k_\perp) \to \gamma(l_\gamma)+ q(l_q)
\end{equation}
where $\bar P^\mu=\bar P^- n^\mu$ and  $ P^\mu= P^+ p^\mu$ with
the usual light cone vectors
$n^\mu$ and $p^\mu$,
normalized according to $p \cdot n=1$. The Mandelstam variables are
defined as: $\hat s=(l_q+l_\gamma)^2$, $\hat t=(xP-l_\gamma)^2$ and $\hat u=(xP-l_q)^2$.
In the correlation limit an effective TMD factorization should apply.
Neglecting the  transverse momentum carried by the incoming quark
the corresponding unpolarized Born cross section reads~\cite{Dominguez:2010xd},
\begin{equation}
\frac{d \sigma}{d P.S.}
= \sum_q x f_q(x)  \, x_g' G_{DP}(x_g',q_{\perp})  H_{Born}
 \,.
\end{equation}
Note that a $\cos 2\phi$ modulation
 will show up for the virtual photon-jet production~\cite{Metz:2011wb}.
The above cross section can also be derived in the color glass condensate(CGC) framework.
 By applying a corresponding power counting in the correlation limit,
a complete matching between  TMD factorization and the CGC calculation
 has been found~\cite{Dominguez:2010xd}.  This power expansion can actually be performed
 either in coordinate~\cite{Dominguez:2010xd} or momentum space~\cite{Akcakaya:2012si}.

For the SSA we proceed in the same way, but include the incoming quark
transverse momentum dependence. To compute the
polarized cross section, one has to apply GTMD factorization as
already without color entanglement
the color flow is non-trivial. This results in,
\begin{eqnarray}
\frac{d \sigma}{d P.S.}
 &=& \sum_q  H_{Born} \int d^2k_{1\perp} d^2k_{2\perp} \delta^2(q_\perp-k_{1\perp}-k_{2\perp})
\nonumber \\
& \times& \!\!\! ~\left \{ x f_q(x,k_{2\perp})  \, x_g'
G_{DP}(x_g',k_{1\perp}) \right.
\nonumber \\
&+& \!\!\! \left. \frac{\varepsilon_\perp^{ij}S_\perp^i k_{2\perp}^j
}{M_N} xf^{\perp,qg \to \gamma q}_{1T,q}(x,k_{2\perp}) x_g'
G_{DP}(x_g',k_{1\perp}) \right \}~~~ \label{4}
\end{eqnarray}
 $f^{\perp,qg \to \gamma q}_{1T,q}(x,k_{2\perp})$
is the quark Sivers function which contains a process dependent gauge link.
Note that the polarized cross section  for photon-jet production in pp collisions takes
the same form~\cite{Bacchetta:2007sz}.
At small $x_g'$, the typical gluon transverse momentum in the nucleus is of order
of the saturation scale $Q_s$, which is
much larger than the intrinsic parton transverse momentum in the proton.
We thus can approximate the cross section by a power expansion in
$k_{2\perp}/k_{1\perp}$
in the semi-hard region where $q_\perp$ is of the order of $Q_s$.
The leading non-trivial contribution reads,
\begin{eqnarray}
&& \!\!\!\!\!\!\!\!\!\!\!\!\!\!\!\!\!\!
  \frac{d \sigma}{d P.S.}
\approx \sum_q H_{Born}
 \left \{
x f_q(x)  \, x_g' G_{DP}(x_g',q_{\perp})\right.
\nonumber \\
&-&
\left.\varepsilon_\perp^{ij}S_\perp^i q_{\perp}^j \frac{N_c^2+1}{N_c^2-1} x T_{F,q}(x,x)
 \frac{\partial}{\partial q_\perp^2}x_g' G_{DP}(x_g',q_{\perp})
\right \}~.
\label{eq7}
\end{eqnarray}
To arrive at the above expression, we have made use of the
identity~\cite{Boer:2003cm,Bomhof:2006ra,Bacchetta:2007sz},
\begin{eqnarray}
\frac{N_c^2+1}{N_c^2-1} T_{F,q}(x,x)=\int d^2k_{2\perp} \frac{k_{2\perp}^2}{M_N}
f^{\perp,qg \to \gamma q}_{1T,q}(x,k_{2\perp})  ~.
\label{eq_7}
\end{eqnarray}
The non-trivial color factor appearing on the left side of Eq.(\ref{eq_7}) is determined by the
color topology of the involved partonic scattering diagram.
It would be interesting to see how the cross section with color entanglement effect being incorporated
deviates from Eq.~\ref{eq7}.

\section{The color entanglement contribution}

Color entanglement effect results from the nontrivial interplay
between gluon attachments from the proton and nucleus sides. In
principle it would be necessary to resum gluon attachments on both
sides to all orders simultaneously. However, such a calculation is
technically out of reach. We simplify the task by taking into
account only one extra collinear gluon from the proton while
resumming gluon re-scattering on the nucleus side to all orders. We
expect this approximation to be valid in the kinematic limit $
q_\perp\sim  Q_s  \sim k_{1\perp} \gg \Lambda_{QCD}\sim k_{2\perp}$
for the following reason. We indicate the transverse momentum
dependence of the Wilson lines by the notation  $\Gamma
(q_\perp-k_{2\perp})$. In the semihard region, one can Taylor expand
this expression $\Gamma (q_\perp-k_{2\perp})=\Gamma (q_\perp)
-2k_{2\perp}\cdot q_\perp \frac{\partial }{\partial q^2_\perp} \Gamma
(q_\perp)+{\cal O}(\frac{k_{2\perp}}{Q_s })$ and conclude that
the $k_{2\perp}^2$ moment of the Sivers function gives rise to the
leading power contribution. To include the $k_{2\perp}^2$  term it
is sufficient to only consider one gluon exchange for the proton
side.

Typically, multiple scattering
between the incoming quark (or transversely polarized gluon) and the classical color field of the
nucleus can be resummed into a Wilson line. However, this
procedure does not apply if the incoming parton is
a collinear gluon.  The formula valid for a longitudinally  polarized gluon scattering off a
nucleus has been worked out in Ref.~\cite{Blaizot:2004wu}. The
expression for  the gauge field created through the fusion of the
incoming gluon from the proton and small x gluons from the nucleus
contains  both singular terms (proportional to $\delta(z^+)$) and regular terms:
$ A^\mu=A^\mu_{reg}+\delta^{\mu-} A^-_{sing} $. The detailed expression for
$A^\mu_{reg}$ and  $A^-_{sing}$ can be found in Ref.~\cite{Blaizot:2004wv}.

\begin{figure}[t]
\begin{center}
\includegraphics[width=9 cm]{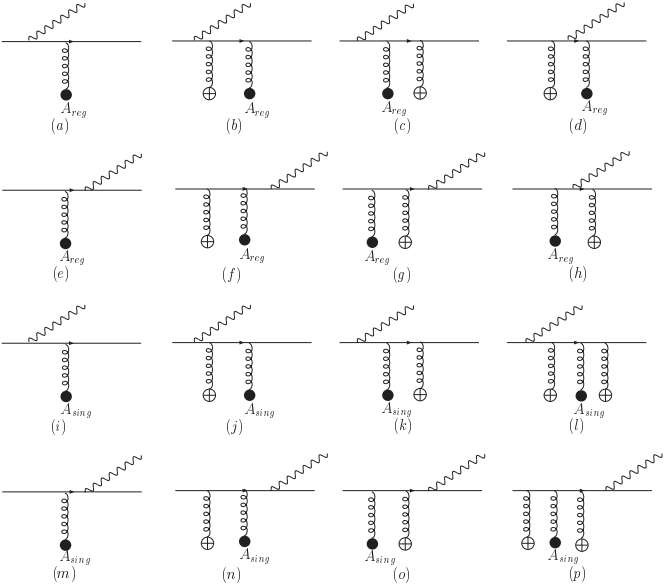}
\caption[] { Diagrams contributing to the spin dependent photon jet production amplitude.
Different symbols indicate different parts of the classical gluon field.
A black dot denotes $A_{reg}$ or $A_{sing}$, while  a cross
surrounded with a circle denotes $A_A$.}
 \label{1}
\end{center}
\end{figure}
We first consider the gluon attachment on the left side of the cut.
All possible insertions of the fields $A_A$, $A^\mu_{reg}$ and  $A^-_{sing} $ on the quark line
must be taken into account as illustrated in Fig.1, where $A_A$ is the classical field created
by the nucleus alone.
As compared to~\cite{Schafer:2014zea},
the calculation of the hard scattering amplitude is greatly simplified
since we do not need to keep track of the transverse momentum carried by
the gluon from the proton side.
The contribution from the initial interactions to the amplitude is given by,
\begin{eqnarray}
&&  \!\!\!\!\!\!\!\!\!\!\!\!\!\!
{\cal M}_I =
 iegA_p^a \int d^2 x_\perp e^{ik_\perp
\cdot x_\perp}   \hskip 2 cm ~~
\nonumber \\
&\times & \!\!\! \bar u(l_q) \frac{  n
\!\!\!/ S_F(xP-l_\gamma) \varepsilon \!\!\!/  + \varepsilon
\!\!\!/S_F(l_q+l_\gamma) n \!\!\!/ }{x_gP+i\epsilon} t^b
U( x_\perp)  u(xP)
\nonumber \\
&\times& \!\!\! \left [ \tilde U(x_\perp)-1 \right ]_{ba}
\end{eqnarray}
where  $A_p^a$ is the gauge field created by the proton  with color index $a$,
and $\varepsilon^\mu$ is the  polarization vector of the produced photon.
$S_F(xP-l_\gamma)=i\frac{xP\!\!\!\!/-l_\gamma\!\!\!\!\!/}{(xP-l_\gamma)^2+i\epsilon}$ and
$S_F(l_q+l_\gamma)=i\frac{l_q\!\!\!\!\!/+l_\gamma\!\!\!\!\!/}{(l_q+l_\gamma)^2+i\epsilon}$ are the quark propagators.
$\tilde U(x_\perp)$ and $U(x_\perp)$ are the path ordered Wilson lines in the adjoint and
fundamental representation
\begin{eqnarray}
\tilde U(x_\perp)&=&{\cal P} {\rm exp} \left [ ig \int_{-\infty}^{+\infty} dz^+ A_A^-(z^+,x_\perp) \cdot T \right ]
 \\
 U(x_\perp)&=& {\cal P} {\rm exp} \left [ ig \int_{-\infty}^{+\infty} dz^+ A_A^-(z^+,x_\perp) \cdot t \right ]
\end{eqnarray}
with $T$  and $t$ being the generators in the adjoint and fundamental representation.
In the case of the SSA for inclusive direct photon production, the contribution from the final state interactions to
the spin asymmetry is absent due to cancelation between mirror diagrams.  However, for the process under consideration,
 the final state interactions shown in diagram (b) and (d) of Fig.1 also generate a spin asymmetry.
 The cancelation  between mirror diagrams does not occur because the out-going quark jet momentum
 is not integrated over.
 The amplitude for final state interactions  reads,
\begin{eqnarray}
&&\!\!\!\!\!\!\!\!
 {\cal M}_F =   iegA_p^a \int d^2 x_\perp e^{ik_\perp
\cdot x_\perp}
 \\ && \!\!\!\!\!\!\!\!
 \bar u(l_q) \frac{  n
\!\!\!/ S_F(xP-l_\gamma) \varepsilon \!\!\!/  + \varepsilon
\!\!\!/S_F(l_q+l_\gamma) n \!\!\!/ }{x_gP-i\epsilon} t^a
  \left [ U( x_\perp)-1 \right ]  u(xP)
\nonumber
   \label{pamp}
\end{eqnarray}
Combining both contributions, we obtain,
\begin{eqnarray}
&& \!\!\!\!\!\!\!\!\!\!\!\!\!\!\!\!
{\cal M}={\cal M}_I+{\cal M}_F
= ieg A_p^a
 \int d^2 x_\perp e^{ik_\perp\cdot x_\perp} \hskip 1 cm
\\
& \times& \!\!\!
 \left \{ {\rm P} \frac{1}{x_gP} \bar u(l_q)   \Big [
n\!\!\!/ S_F(xP-l_\gamma) \varepsilon \!\!\! / + \varepsilon
\!\!\!/S_F(l_q+l_\gamma) n \!\!\!/ \Big ] \right.
\nonumber\\
&& \ \ \ \times t^b
   U( x_\perp)  u(xP)\left [ \tilde U(x_\perp) \right ]_{ba}
\nonumber \\
&&-i\pi\delta(x_gP)  \bar u(l_q)   \Bigl[
n\!\!\!/ S_F(xP-l_\gamma) \varepsilon \!\!\!/
+ \varepsilon
\!\!\!/S_F(l_q+l_\gamma) n \!\!\!/ \Bigr]
\nonumber \\
&& \left. \ \ \ \times t^b
   U( x_\perp)  u(xP)\left [ \tilde U(x_\perp) -2 \right ]_{ba} \right \}
\nonumber
\label{amp}
\end{eqnarray}
where a term which does not contain any Wilson line has been neglected. The contribution from such a term
to the inelastic scattering cross section is suppressed according to the arguments made in Ref.~\cite{Dumitru:2002qt}.
Using the identity $\frac{1}{x_gP \pm i \epsilon}=P \frac{1}{x_gP}\mp i\pi \delta(x_gP)$,
the real and imaginary part of the gluonic pole have been expressed separately in the above equation.
The imaginary part provides the phase necessary
for generating a non-vanishing SSA, whereas the real part is irrelevant for the polarized cross section.
Note that the structures of the Wilson lines associated with the real and imaginary parts are different.
After carrying out the  $x_g$ integration,
the gluon field $A_p^a$ is incorporated into the gauge links of the unpolarized quark TMD and the quark Sivers function.

At the order we consider, the spin dependent hard part is calculated from an interference of the scattering amplitude with
 one extra gluon attachment from the proton given in the above equation
and the Born scattering amplitude derived in Ref.~\cite{Gelis:2002ki}.
 It is straightforward to include the contributions from the right cut diagrams.
To arrive at a compact expression for the polarized cross section, we simplify the Wilson line structure using the
approach introduced in~\cite{Schafer:2014zea}. As a result, two different Wilson line structures emerge.
These can be related to the dipole type gluon TMD distribution, and a
new gluon distribution $G_4$, respectively.
We then proceed by extrapolating the result to the correlation limit
 with a power expansion procedure performed in momentum space~\cite{Akcakaya:2012si}.
 After neglecting all terms suppressed by the power of  $k_{1\perp}/l_{q\perp}$,
 the hard part is no longer dependent of the incoming parton transverse momenta,
 while the soft part is expressed as the convolution of the quark Sivers function and the gluon distributions $G_{DP}$ and $G_4$.
As argued above, the single gluon exchange from the proton remains a good approximation in the semihard region.
In this kinematic region, the soft part can be further simplified by carrying out power expansion leading from
Eq.(\ref{4}) to Eq.(\ref{eq7}).
 Eventually, one obtains  Eq.(\ref{eq_1}).

The definition of  $G_4(x_g',k_\perp)$ has been given in~\cite{Schafer:2014zea},
\begin{eqnarray}
x_g' G_{4}(x_g', k_\perp)&=& \frac{k_\perp^2 N_c}{2 \pi^2 \alpha_s}
\int \frac{d^2x_\perp d^2y_\perp }{(2\pi)^2} e^{i  k_\perp \cdot
(x_\perp-y_\perp)}
\nonumber\\ & \times&
\frac{1}{N_c^2} \langle {\rm Tr_c} [U(x_\perp)]
{\rm Tr_c}[ U^\dag(y_\perp)] \rangle_{x_g'}
\label{eq_a}
\end{eqnarray}
It is important to note that the extra gluon attachment from the
proton plays a crucial  role in yielding the non-trivial Wilson line
structure appearing in Eq.(\ref{eq_a}). The both distributions
$G_4(x_g',k_\perp)$ and $G_{DP}(x_g',k_\perp)$ can be evaluated
within the MV model,
\begin{eqnarray}
x_g' G_{4}(x_g', k_\perp)&=&\frac{k_\perp^2
R_0^2}{2 \pi \alpha_s N_c}
\int \frac{d^2r_\perp }{(2\pi)^2} e^{i  k_\perp \cdot r_\perp}
 e^{-\frac{1}{4} r_\perp^2
Q_{sq}^2}  \ \nonumber\\ & =&
\frac{1}{N_c^2} x_g'G_{DP}(x_g', k_\perp)
\end{eqnarray}
where $R_0$ is the radius of the nucleus. One notices that  $G_4$ is suppressed in the
large $N_c$ limit as compared to the normal dipole type gluon distribution $G_{DP}$.

We close this section with two further
remarks:\\
(i) The unpolarized cross section is not affected by the color entanglement effect at the order that we consider.
The observed color entanglement effect is the consequence of the non-trivial interplay
among the T-odd effect, the coherent multiple gluon re-scattering, and the non-Abelin feature of QCD.
\\
(ii) The spin asymmetry is found to be proportional to the slope of
both gluon distributions. The same feature was also found
in~\cite{Boer:2006rj} for a different process.

\section{conclusions}
We have shown that the SSA in photon-jet production in polarized pA collisions in the correlation limit
offers a promising opportunity to pin down experimentally the size of the factorization breaking
color entanglement effect parameterized by the small x gluon distribution $G_4$.
We hope very much that such a measurement will be performed at RHIC~\cite{Aschenauer:2013woa}.

\vskip 1 cm {\bf Acknowledgments:} We are grateful to Bowen Xiao and
Feng Yuan for reminding us of  the tadpole type contribution to
  the gluon distribution $G_4$ in the MV model.
 This work has been supported by BMBF (OR 06RY9191).

\end {document}